# Tunable Klein-like tunneling of high-temperature superconducting pairs into graphene

David Perconte[1,*], Fabian A. Cuellar[1,*], Constance Moreau-Luchaire[1], Maelis Piquemal-Banci[1], Regina Galceran[1], Piran R. Kidambi[2], Marie-Blandine Martin[2], Stephan Hofmann[2], Rozenn Bernard[1], Bruno Dlubak[1], Pierre Seneor[1,♣] and Javier E. Villegas[1,♦]

[1] *Unité Mixte de Physique, CNRS, Thales, Univ. Paris-Sud, Université Paris Saclay, 91767 Palaiseau, France*
[2] *Department of Engineering, University of Cambridge, Cambridge CB3 0FA, United Kingdom*

[*] these authors equally contributed
[♣] pierre.seneor@cnrs-thales.fr
[♦] javier.villegas@cnrs-thales.fr

**Superconductivity can be induced in a normal material via the "leakage" of superconducting pairs of charge carriers from an adjacent superconductor. This so-called proximity effect is markedly influenced by graphene's unique electronic structure, both in fundamental and technologically relevant ways. These include an unconventional form[1,2] of the "leakage" mechanism –the Andreev reflection[3]– and the potential of supercurrent modulation through electrical gating[4]. Despite the interest of high-temperature superconductors in that context[5,6], realizations have been exclusively based on low-temperature ones. Here we demonstrate gate-tunable, high-temperature superconducting proximity effect in graphene. Notably, gating effects result from the perfect transmission of superconducting pairs across an energy barrier –a form of Klein tunneling[7,8], up to now observed only for non-superconducting carriers[9,10]– and quantum interferences controlled by graphene doping. Interestingly, we find that this type of interferences become dominant without the need of ultra-clean graphene, in stark contrast to the case of low-temperature superconductors[11]. These results pave the way to a new class of tunable, high-temperature Josephson devices based on large-scale graphene.**

Superconductivity is induced in a normal metal (N) in contact with a superconductor (S) via the Andreev reflection(AR)[3]: an electron entering S from N pairs to another electron to form a Cooper pair, leaving a hole-like quasiparticle that is transmitted back into N. Electron and hole coherently propagate with parallel opposite wave-vectors, carrying superconducting correlations into N. This mechanism allows supercurrent flow and Josephson coupling across S-N-S junctions[12].

S-N proximity devices that can be greatly tune by electrostatic doping are one of the main technological prospects of induced superconductivity in graphene[4,13–16]. Several specific mechanisms allow for that. Besides the density-of-states narrowing at the Dirac point, around which the junction's resistance increases, subtler effects may play a role. For example, the unusual possibility that AR involved electron and hole reside in different bands –conduction and valence– results in a Specular Andreev Reflection[1] (SAR) in which electron and hole wave-vectors are mirror-like. SAR can occur if the graphene's Fermi energy $E_F$ is lower than the superconducting energy-gap $\Delta$, while for $E_F > \Delta$ the conventional (intra-band) AR takes place[1]. Thus, an AR to SAR crossover can be driven by shifting $E_F$ through a gate voltage, which dramatically changes the S-N interface conductance[2].

In the present experiments, tunability results from a different mechanism that involves Klein tunneling –i.e. the reflectionless transmission of electrons across a high energy barrier[7–10]. Beyond single electrons, here we observe Klein-like tunneling of Andreev electron-hole pairs that carry superconducting correlations from the high-temperature superconductor $YBa_2Cu_3O_7$ (YBCO) into graphene. That effect is associated to quantum interferences across the barrier, which modulate the overall penetration of superconducting pairs into graphene. Those interferences are gate-tunable because the wave-vector of graphene's massless charge carriers is proportional to $E_F$. While extremely sensitive to $E_F$ inhomogeneities comparable to $\Delta$

–which has confined their pristine observation to ultraclean exfoliated graphene[11] – we find that the large Δ of YBCO makes interference effects robust in Chemical Vapor Deposited (CVD) graphene. This increases the technological potential of the observed effects.

Fabricating YBCO/graphene devices with electron-transparent interfaces had remained challenging[17]. Contrary to low-temperature superconductors[2,4,11,13–16,18–20], YBCO cannot be grown on graphene due to its deposition conditions (hundreds of °C, oxygen-rich atmosphere). Conversely, the surface electronic properties of YBCO are easily degraded by standard graphene fabrication and lithography techniques. To circumvent those constraints, we used 50 nm thick YBCO films grown on SrTiO$_3$ (STO) and covered it *in situ* with an ultrathin 4 nm Au layer. This protects the YBCO surface and constitutes a high-transparency interface with graphene. Then we applied a masked ion irradiation technique (details in Fig.1) that allowed us to fabricate planar devices as shown in Fig. 1e: four superconducting YBCO/Au electrodes (bright color) defined within an insulating YBCO matrix (dark color) are bridged by an overlaying single-layer CVD graphene sheet. With this device, we study the YBCO/Au/graphene junction conductance using a three-probe configuration, with the current $I$ injected between electrodes 1 and 4 and the voltage $V$ measured between 3 and 4 [sketch in Fig. 1 (e)]. In that configuration the contributions of the YBCO leads and the YBCO/Au interface to the measured conductance are negligible (see Supplementary Information S1). Thus, the measurement allows us to probe the Au/graphene conductance and, particularly, the proximity behavior of that *single* interface. Notice also that the ultrathin Au layer is expected to sustain proximity-induced d-wave superconductivity (see Supplementary Information S1). We stress that here we do not investigate Josephson effect across graphene, which is not observed nor expected because the distance between YBCO/Au electrodes (~5 μm) is much larger than the estimated coherence length in graphene $\xi_G = \sqrt{\hbar D/k_B T}$ ~ 270 nm at 4K. Back-gating through the dielectric STO substrate (scheme in Fig. 2c) is used to investigate graphene doping effects on the proximity behavior.

Fig. 2a shows the differential conductance $G = dI/dV$ vs. $V$ at several temperatures T in zero applied magnetic field. For T well above the critical temperature $T_C \sim 80K$, the $V$ dependence is relatively weak and $G$ decreases with increasing $|V|$. Upon decreasing T below $T_C$, a large conductance enhancement develops around zero-bias, accompanied by two dips near ±20 mV. Those features become more pronounced as T decreases. At the lowest T, the zero-bias conductance is nearly twice the conductance at $V$=60 mV (this is well above the YBCO superconducting gap[21]). Magnetic fields up to up to $H$=6 kOe had no effect on $G(V)$ (data not shown). In contrast, $G(V)$ is strongly modulated by back-gating.

Following the scheme in Fig. 2c, the application of a gate voltage $V_G$ leads to graphene doping due to the polarization of the substrate (STO) and the irradiated (insulating) YBCO. Fig. 2b illustrates $V_G$ effects. For convenience, the normalized conductance $g(V) \equiv G(V)/G_N(V)$ is shown, with $G_N(V)$ the normal-state one at 120 K. The conductance under $V_G = -40$ V (bottom panel) is markedly different to that for $V_G = 0$ V (top panel): changes in the zero-bias peak width and height, in the dips' depth and in the conductance background are observed. Notice e.g. a conductance enhancement ($g > 1$) for $V_G = -40$ V (bottom) within the bias range in which dips and $g < 1$ are observed for $V_G = 0$ V (top). As discussed below, the curves in Fig. 2b constitute two behavior types between which $g(V)$ periodically switches as a function of $V_G$.

Fig. 3a displays $g(V, V_G)$ in a contour plot generated from a series of measurements as those in Fig. 2b (data range is $|V|$< 30 mV (around 1.5Δ) and $|V_G|$ <100 V in this figure; the data for the whole experimental window is displayed in Fig. S6). The zero-bias peak observed for all $V_G$ (Fig. 2b) appears in the contour plot as a vertical "band" (in red). The conductance dips observed around ± ~20 mV in Fig. 2b show in the plot as "pockets" (in purple color). The background enhancement ($g > 1$) seen in Fig. 2b for $V_G = -40$ V and $|V| < 40$ mV shows in the plot as a horizontal feature (in green). Purple "pockets" and green horizontal features alternate periodically along the y-axis, i.e. as a function of $V_G$. This periodic modulation is highlighted in Fig. 3b, which displays $g(V_G)$ for fixed $V$ –this corresponds to vertical "cuts" of the contour plot, marked with dashed lines Θ and Σ in Fig. 3a. Notice (Fig. 3b) that the periodic modulation is accompanied by a conductance background decrease for increasing $V_G$. Further analysis of the oscillatory behavior can be found in Figure S7.

In order to interpret the experimental results, we start by considering the conductance across generic superconductor/normal (S/N) interfaces[22]. In this context, the zero-bias conductance doubling at low-T (Fig. 2a) suggests a highly transparent S/N interface governed by AR[22] (devices showing reduced transparency are discussed in the Supplementary Information S2). The invariance of the conductance curves with respect to the magnetic field is consistent with that scenario. $V_G$ effects imply that graphene is necessarily the "N" part of the S/N junction, and rule out a possible role of the YBCO/Au interface[23]. This is because the device geometry ensures that $V_G$ solely affects graphene that lies on insulating YBCO (dark in Fig. 2c). The electric field is fully screened elsewhere by the superconducting YBCO (bright in Fig. 2c, whose properties are unaffected by $V_G$ because its thickness is ~50 times the Thomas-Fermi screening length[24].

The overall experimental details are explained by the model sketched in Figs. 3g. This model is based on the expectation that graphene on top of superconducting YBCO/Au becomes superconducting (S) by proximity effect[1] and presents a fixed doping $E'_F$, while graphene on top of insulating YBCO is normal (N) and presents a gate-tunable doping $E_F(V_G)$. Then we assume that there is an intermediate region N' of width $w$ in which graphene is normal but presents different doping than in S and N (see Supplementary Information S5). This scenario leads to a potential energy step $U_0$ for electrons at the boundary between S and N graphene[6] (note that replacing S by normal graphene N results in the type of graphene N-N'-N junction in which regular –non-superconducting– Klein-tunneling is expected[7–10]). The conductance across such an S-N'-N junction was theoretically studied both for *s*-wave[5,6,25] and *d*-wave superconductors[5,6] (note that in those works exactly the same junction structure was referred to as S-I-N). The central effect arises due to interferences of electrons and holes traveling from the N'-N to the S-N' interface (and *vice versa* after reflection at S-N'). Depending on the phase $\chi = -wk_F$ picked up by electrons/holes across N' (with $k_F$ the Fermi vector), constructive or destructive interference occurs, which periodically modulates the junction conductance as a function of $\chi$.[6] Thus, due to graphene's linear dispersion, $\chi = w(E_F - U_0)/\hbar v_F$ (with $v_F$ the Fermi velocity) and a periodic conductance modulation as a function of $E_F$ is expected. That can explain the gating effects observed in Fig. 3a and b. To support this, we used the existing theory[6] to perform the numerical simulations shown in Fig. 3d-f.

Simulations of $g(eV/\Delta, \chi)$ require several input parameters: i) the native Fermi energy in N, $E_F(0)$; ii) the Fermi energy in S, $E'_F$; ; iii) the proportionality factor between $E_F$ variation and phase shift $\delta E_F/\delta\chi$ ; and iv) the angle $\alpha$ between the *d*-wave nodes and the S-N' interface[5]. As discussed in the Supplementary Information S4, most of these parameters are known from independent experiments, and only $\alpha$ and $E'_F$ remain as fitting parameters. Fig. 3d shows the best agreement between theory and experiments, obtained for $\alpha = \pi/4$, $E'_F = 20\Delta$, $E_F(0) = 17.5\Delta$, and $\delta E_F/\delta\chi = 5\Delta/3\pi$. Notably, all of the characteristic features seen in the experimental $g(V, V_G)$ (Fig. 3a) can be found in Fig. 3d: a vertical band (in red) that corresponds to zero-bias conductance peak due to AR, and a periodic modulation along the y-axis in which horizontal features due to enhanced conductance (in green) alternate with "pockets" (in purple). Figure 3e show vertical "cuts" of the theoretical plot along the lines Θ and Σ, which closely mimic the experimental ones (Fig. 3b). "Cuts" along the x-axis (labeled A, B, C, D) are also shown for experiments and theory, respectively in Fig. 3c and 3f. Their comparison show that the simulations not only successfully reproduce the periodic gating effects, but also the essential details of the experimental $g(V)$: a conductance enhancement around zero bias, "dips" around eV~Δ, and the evolution of these features with $V_G$ (phase χ). The overall agreement between simulation and experiment demonstrates that the used model[6] captures the physics of the studied system. Further improving the match between theory and experiment, particularly for eV> ~1.5Δ where discrepancies appear (see Fig. S6), may require refining the model, e.g. by considering different barrier profiles (e.g. trapezoidal instead of square), smeared or rough S/N'/N interfaces, etc (see discussion in Supplementary Information section 5). We hope that the present experiments will motivate that work.

In summary, the periodic modulation of the conductance is explained by electron interferences within an energy barrier that separates normal and superconducting graphene. The barrier is essentially transparent to electron/hole superconducting pairs (as well as for normal electrons) when the phase picked across the barrier is $\chi = n\pi$, which corresponds to Klein tunneling of superconducting pairs. Thus, the interferences allow for an electrostatic tuning of the superconducting proximity effect, since the fraction of the current carried by electron-hole superconducting pairs (created by AR) is modulated by $\chi(V_G)$[6]. These results open interesting perspectives. For instance, experiments aimed at selecting the pairing symmetry induced in graphene via the control of the orientation of the graphene and YBCO lattices[26], gate-tunable high-temperature Josephson devices using large-scale –not necessarily ultraclean– graphene, and possibly more elaborate devices in which those effects, the salient features of *d*-wave superconductivity and the directional nature of Klein tunneling[7] may be exploited altogether, e.g. to create tunable "pi" Josephson junctions[27].

**Note added:** during the resubmission of the paper we learnt about the publication of related Scanning Tunneling Microscopy experiments of graphene on a different cuprate superconductor [28].

**Acknowledgement**s: Work at CNRS/Thales supported by the French National Research Agency through "Investissements d'Avenir" program Labex NanoSaclay (ANR-10-LABX-0035) and by the EU Work Programme under Grant Graphene Flagship (No. 604391) and Core1 (No. 696656). P.S. acknowledges the Institut Universitaire de France for a junior fellowship. We thank A. S. Mel'Nikov, J. Linder, J. Santamaría, S. Gueron and H. Bouchiat for useful discussions. We thank Y. Le Gall for assistance during ion irradiation.

**Methods**

c-axis YBCO thin films (50 nm thick) were grown on (100) SrTiO$_3$ (STO) substrates (500 µm thick) by pulsed-laser deposition, at 700°C and 0.35mbar of pure oxygen. The optimum oxygen stoichiometry is ensured by cooling-down to room temperature in 800 mbar of pure oxygen. Subsequently the chamber base pressure 3×10$^{-6}$ torr is reinstated, and an ultra-thin Au film (~4 nm thick) is deposited *in situ* on top of the YBCO to protect its surface. YBCO c-axis epitaxial growth is confirmed by RHEED (*in situ*) and X-ray diffraction (*ex situ*), and film thicknesses are determined by X-ray reflectivity. Atomic Force Microscopy of the Au/YBCO bilayer shows flat surfaces with typical rms roughness ~3 nm.

Graphene transfer onto the lithographed devices was done as reported elsewhere.[29]

Electrical characterization within the 3K-300K range was carried in a He-flow cryostat equipped with a 6kOe electromagnet. In the 3-probe configuration, the current is injected between contacts 1 and 4, and the voltage probes are attached to contacts 3 and 4 (see image in Fig. 1e). dI/dV (V) is obtained using the current-biased Keithley delta-mode® with coupled K2182 nanovoltmeter and K6221 current source.

**Data availability.**

The data supporting the plots within this paper and other findings of this study are available from the corresponding author on reasonable request.

1. Beenakker, C. W. J. Specular andreev reflection in graphene. *Phys. Rev. Lett.* **97,** 67007 (2006).
2. Efetov, D. K. *et al.* Specular interband Andreev reflections at van der Waals interfaces between graphene and NbSe2. *Nat. Phys.* **12,** 328–332 (2015).
3. Andreev, A. F. The thermal conductivity of the intermediate state in superconductors. *Sov Phys JETP* **19,** 1228–1231 (1964).
4. Heersche, H. B., Jarillo-Herrero, P., Oostinga, J. B., Vandersypen, L. M. K. & Morpurgo, A. F. Bipolar supercurrent in graphene. *Nature* **446,** 56–9 (2007).
5. Linder, J. & Sudbø, A. Dirac fermions and conductance oscillations in s- and d-wave superconductor-graphene junctions. *Phys. Rev. Lett.* **99,** 147001 (2007).
6. Linder, J. & Sudbø, A. Tunneling conductance in s- and d-wave superconductor-graphene junctions: Extended Blonder-Tinkham-Klapwijk formalism. *Phys. Rev. B* **77,** 64507 (2008).
7. Katsnelson, M., Novoselov, K., & Geim, A. Chiral tunnelling and the Klein paradox in graphene. *Nat. Phys.* **2,** 620–625 (2006).
8. Beenakker, C. W. J. Colloquium: Andreev reflection and Klein tunneling in graphene. *Rev. Mod. Phys.* **80,** 1337–1354 (2008).
9. Huard, B., Sulpizio, J. A., Stander, N., Todd, K., Yang, B. & Goldhaber-Gordon, D. Transport Measurements Across a Tunable Potential Barrier in Graphene. Physical Review Letters, 98(23), **236803,** 8–11 (2007).
10. Young, A. F. & Kim, P. Quantum interference and Klein tunnelling in graphene heterojunctions. *Nat. Phys.* **5,** 222–226 (2009).
11. Ben Shalom, M. *et al.* Quantum oscillations of the critical current and high-field superconducting proximity in ballistic graphene. *Nat. Phys.* **12,** 318–322 (2015).
12. Klapwijk, T. M. Proximity Effect From an Andreev Perspective. *J. Supercond.* **17,** 593–611 (2004).
13. Du, X., Skachko, I. & Andrei, E. Y. Josephson current and multiple Andreev reflections in graphene SNS junctions. *Phys. Rev. B.* **77,** 184507 (2008).
14. Ojeda-Aristizabal, C., Ferrier, M., Guéron, S. & Bouchiat, H. Tuning the proximity effect in a superconductor-graphene-superconductor junction. *Phys. Rev. B* **79,** 165436 (2009).
15. Girit, Ç. *et al.* Tunable graphene dc superconducting quantum interference device. *Nano Lett.* **9,** 198–199 (2009).
16. Komatsu, K., Li, C., Autier-Laurent, S., Bouchiat, H. & Guéron, S. Superconducting proximity effect in long superconductor/graphene/superconductor junctions: From specular Andreev reflection at zero field to the quantum Hall regime. *Phys. Rev. B* **86,** 115412 (2012).
17. Sun, Q. J. *et al.* Electronic transport transition at graphene/YBa2Cu3O7−δ junction. *Appl. Phys. Lett.* **104,** 102602 (2014).
18. Rickhaus, P., Weiss, M., Marot, L. & Schönenberger, C. Quantum hall effect in graphene with superconducting electrodes. *Nano Lett.* **12,** 1942–1945 (2012).
19. Deon, F., Šopić, S. & Morpurgo, A. F. Tuning the influence of microscopic decoherence on the superconducting proximity effect in a graphene Andreev interferometer. *Phys. Rev. Lett.* **112,** 126803 (2013).
20. Allen, M. T. *et al.* Spatially resolved edge currents and guided-wave electronic states in graphene. *Nat. Phys.* **12,** 128–133 (2016).
21. Wei, J. Y. T., Yeh, N.-C., Garrigus, D. F. & Strasik, M. Directional Tunneling and Andreev Reflection on YBa2Cu3O7−δ Single Crystals: Predominance of d-Wave

Pairing Symmetry Verified with the Generalized Blonder, Tinkham, and Klapwijk Theory. *Phys. Rev. Lett.* **81,** 2542–2545 (1998).
22. Blonder, G. E., Tinkham, M. & Klapwijk, T. M. Transition From Metallic To Tunneling Regimes in Superconducting Micro-Constrictions - Excess Current, Charge Imbalance, and Super-Current Conversion. *Phys. Rev. B* **25,** 4515 (1982).
23. Kashiwaya, S., Tanaka, Y., Koyanagi, M., Takashima, H. & Kajimura, K. Origin of zero-bias conductance peaks in high-Tc superconductors. *Phys. Rev. B* **51,** 1350 (1995).
24. Crassous, A. *et al.* Nanoscale electrostatic manipulation of magnetic flux quanta in ferroelectric/superconductor BiFeO 3/YBa 2Cu 3O 7-δ heterostructures. *Phys. Rev. Lett.* **107,** 247002 (2011).
25. Bhattacharjee, S. & Sengupta, K. Tunneling conductance of graphene NIS junctions. *Phys. Rev. Lett.* **97,** 217001 (2006).
26. Linder, J., Black-Schaffer, A. M., Yokoyama, T., Doniach, S. & Sudbø, A. Josephson current in graphene : Role of unconventional pairing symmetries. *Physical Review B*, **80(9)**, 094522 1–15 (2009).
27. Cedergren, K. *et al.* Interplay between Static and Dynamic Properties of Semifluxons in YBa_{2}Cu_{3}O_{7-δ} 0-π Josephson Junctions. *Phys. Rev. Lett.* **104,** 177003 (2010).
28. Bernard, A. Di *et al.* p-wave triggered superconductivity in single-layer graphene on an electron-doped oxide superconductor. *Nat. Commun.* **8,** 1–9 (2017).
29. Kidambi, P. R. *et al.* The Parameter Space of Graphene CVD on Polycrystalline Cu. *J. Phys. Chem. C* **116,** 22492–22501 (2012).

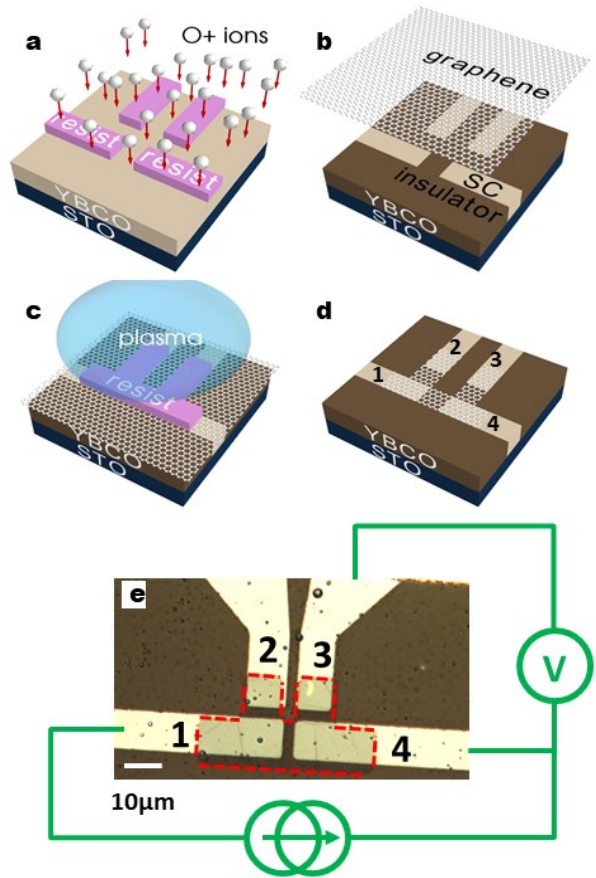

**Figure 1: Sample fabrication steps. (a)** Starting with an Au (4 nm)/YBCO (50 nm) film grown on SrTiO$_3$, photo-lithography is used to define a resist pattern through which the film is irradiated with 110 KeV O$^+$ ions at high fluence (5 10$^{14}$ cm$^{-2}$). This turns the unprotected YBCO electrically insulating. Conversely, the resist-covered YBCO maintains superconducting (S) properties. **(b)** After subsequently removing Au (via Ar$^+$ ion etching) and the resist mask, a planar device is obtained with S electrodes (bright color) defined within an insulating YBCO matrix (dark). A CVD single-layer graphene sheet is then transferred as reported elsewhere[26] **(c)** Graphene is patterned using photo-lithography and oxygen plasma to obtain a channel that bridges the S electrodes. **(d)** Sketch of the final device **(e)** Microscope image of an actual device. The graphene layer has been outlined (red-dashed line) for clarity. In green colors, a sketch of the measurement geometry.

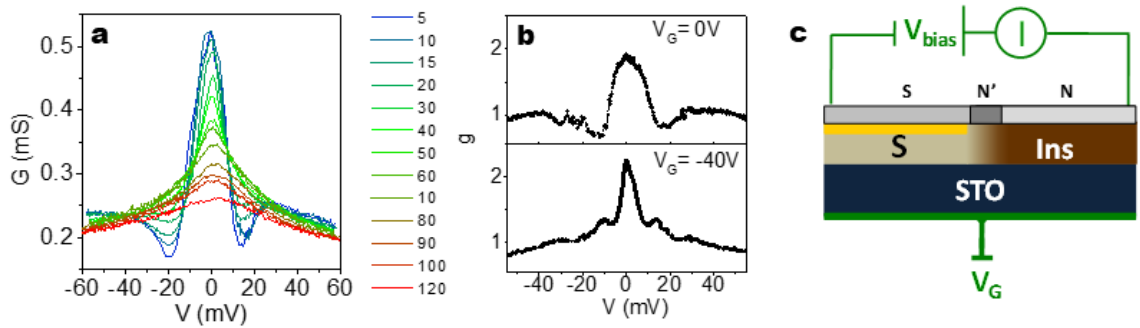

**Figure 2: Temperature and gate-voltage dependent conductance. (a)** Differential conductance of the YBCO/Au/graphene junction in zero applied field and different temperatures (see legend in K). **(b)** Normalized conductance at zero magnetic field and T= 4 K for two gate voltages: $V_G$= 0 V (top) and $V_G$ =-40 V (bottom). **(c)** Scheme of the device cross-section. Graphene, in grey, is S above the superconducting YBCO (light color), and normal (N' and N) elsewhere. The insulating (Ins) YBCO is depicted in dark color. The boundary between superconducting and insulating YBCO has a width of a few tens of nm (degraded color). Gating is achieved across the dielectric STO substrate.

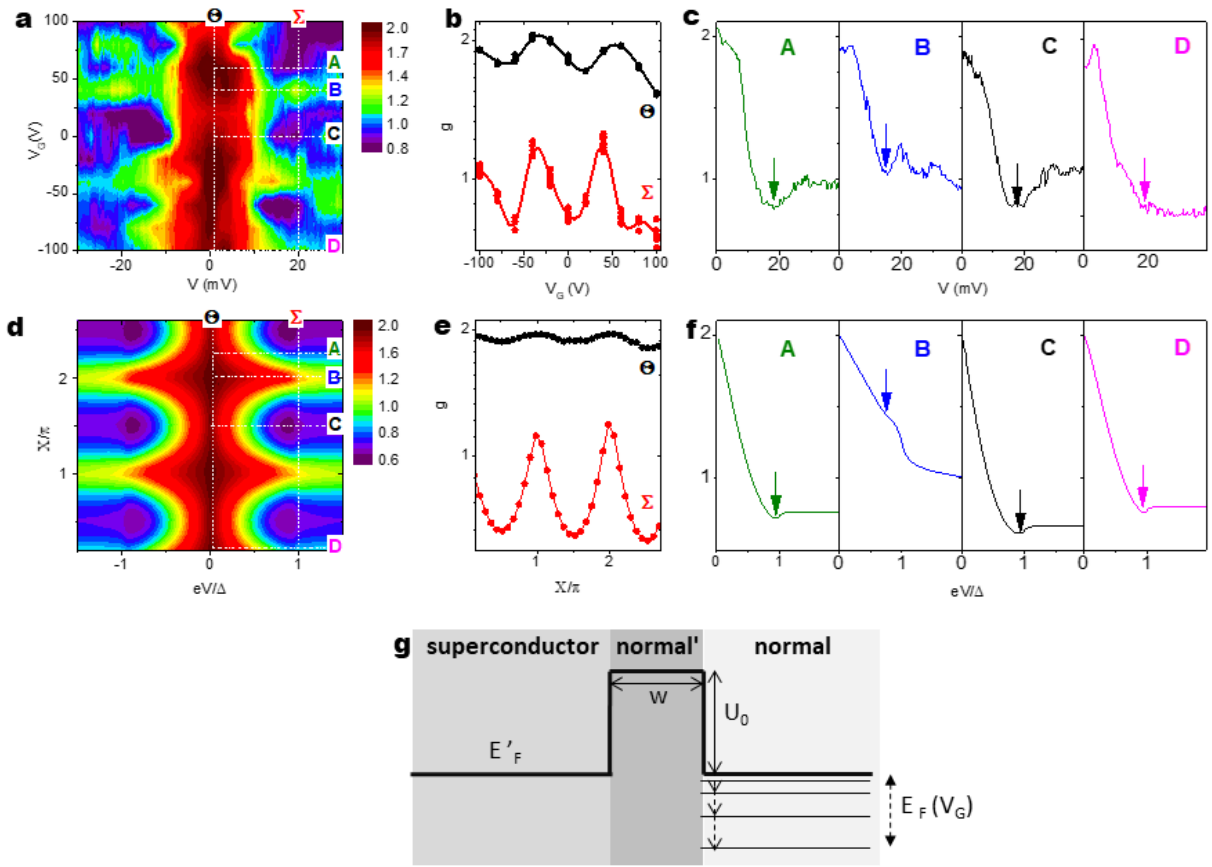

**Figure 3: Gate voltage effects: experiment and simulations. (a)** Contour plot of the normalized conductance (color scale) as a function of junction bias V and gate voltage $V_G$. This plot is constructed from a series of measurements as those shown in Figure 2, with $\Delta V_G$=20 V. **(b)** Experimental conductance a function of gate voltage $V_G$ at fixed bias V, which correspond to vertical "cuts" of the plot in (a) along the dashed lines labeled Θ and Σ (respectively for V=0±1mV and V=20±1 mV). Various points per $V_G$ exists which denote the measurement uncertainty (see Supplementary Information section 7 for further details). The solid line behind the data is a guide to the eye. **(c)** Experimental conductance *vs.* bias V for fixed gate voltages $V_G$. Each panel corresponds to a horizontal "cut" of the plot in (a) along the lines labeled A,B,C,D (respectively for $V_G$= 60, 40, 0 and -100 V). The arrow points to the conductance dip below eV= Δ **(d)**, **(e)** and **(f)** show numerical simulations correlative of the measurements in (a)-(c). Notice that in the simulation the bias V is normalized by the superconducting gap $\Delta$, and the experimental quantity $V_G$ is replaced by the phase X. **(g)** Schematic of the theoretical model. A potential step of width w and height $U_0$ defines a region of normal graphene (N') that separates superconducting graphene with Fermi energy $E'_F$ from normal graphene (N) with gate-tunable Fermi energy $E_F(V_G)$.

Supplementary Information for

Tunable Klein-like tunneling of high-temperature superconducting pairs into graphene

by

David Perconte, Fabian A. Cuellar, Constance Moreau-Luchaire, Maelis Piquemal-Banci, Regina Galceran, Piran R. Kidambi, Marie-Blandine Martin, Stephan Hofmann , Rozenn Bernard, Bruno Dlubak, Pierre Seneor and Javier E. Villegas

1. **Contribution of the YBCO/Au interface and YBCO leads to the measured conductance**

The YBCO/Au contact resistance was measured in a series of YBCO/Au micro-junctions (area 10 to 100 $\mu m^2$) fabricated as reported elsewhere [1] using Au/YBCO bilayers as those used for graphene devices. Note that Au was deposited *in situ* right after the YBCO growth. The contact resistance is ohmic (see Fig. S1), and the contact resistivity ranges in most cases between ~ $10^{-7}$-$10^{-6}$ Ohm $cm^2$ at room temperature. Since Au covers YBCO over a large area (YBCO/Au leads are macroscopic), the contribution of that interface to the measurement is negligible. Even a very conservative estimate, made by assuming instead a $10^2$ $\mu m^2$ Au/YBCO contact area (this roughly corresponds to the overlap with graphene) yields $G$~ 1 – 10 S. This is four to five orders-of-magnitude higher than the measured conductance ~ 0.1 – 1 mS (see Fig. 2).

The contribution of the YBCO leads to the measurement is ruled out because the current circulating through the device is very low, of the order of $I=VG \sim 10^{-2} V \times 10^{-4}$ S ~ 1 µA. The current density across the YBCO leads is therefore ~ $2 \cdot 10^2$ A cm$^{-2}$ at most, as obtained from the YBCO minimal section S= 50 nm (lead thickness) × 10 µm (lead width). That current density is orders-of-magnitude below the YBCO critical current in thin films, which is ~ $10^6$ A cm$^{-2}$ at 77 K. Furthermore, the invariance of the conductance measurements under applied magnetic field and the fact that the high-bias (V>50 mV) conductance is nearly constant above and below Tc (Fig. 2a) clearly demonstrate that YBCO leads are not contributing to the measurement.

The Au layer plays a crucial role. We fabricated devices in which graphene was directly deposited on YBCO, with no Au interlayer. In this case, the conductance was extremely low, leading in most cases to non-measurable, open-circuit devices. This is actually as expected because, when exposed to air and lithography chemicals, the YBCO surface rapidly degrades and becomes insulating. The *in situ* deposition of Au avoids this.

Note finally that, considering earlier experiments and theory, the 4-nm Au interlayer is thin enough to be fully superconducting with a d-wave order parameter:

1) <u>Experimentally,</u> STM measurements on polycrystalline Au deposited on c-axis YBCO showed that the penetration length of superconductivity into Au is ~30 nm (see e.g. [2]). Since in our experiment the Au layer thickness is one order-of-magnitude smaller (4 nm), we do not expect a strong suppression of the induced order parameter.

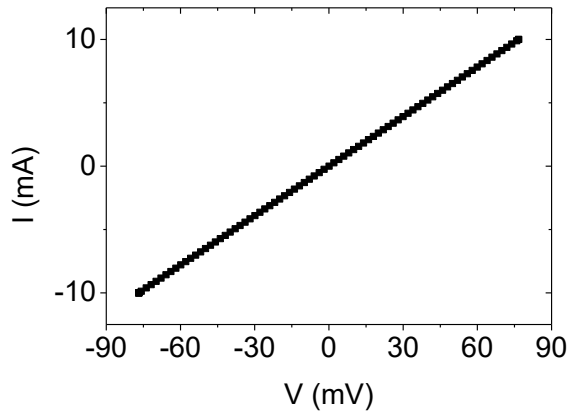

**Figure S1:** Typical I-V characteristic of Au/YBCO contacts.

2) <u>Theoretical studies</u> on the proximity effect between d-wave superconductors and diffusive normal-metals show that, when the c-axis is perpendicular to the superconductor/metals interface (our experiment), the condensate penetrating the normal-metal decays and preserves the d-wave symmetry over the length scale of the normal-metal electronic mean free path (see e.g. [3]). The mean free path in polycrystalline Au is typically of the order of tens of nm [see e.g. [4]], that is, one order-of-magnitude longer that the thickness of the Au layer in our experiments. From this, we expect that the d-wave symmetry is preserved at the gold-graphene interface.

## 2. Effect of the Au/graphene interface transparency

We measured ~20 three-terminal graphene/Au/YBCO devices, whose conductance showed in all cases superconducting-gap related features that were unaffected by applied magnetic field |H| < 6 kOe. Two different types of behaviors were observed. The first one corresponds to that in Figs. 2 and 3 in the main text and in Figure S2 (a), for which the hallmark is a doubling of the conductance around zero bias. Other devices showed instead a moderate conductance decrease with a dip around zero-bias [see examples in Figure S2 (b) and (c)]. The conductance doubling for sub-gap energies suggests a highly transparent S/N interface across which transport is governed by Andreev reflection. Conversely, the moderate conductance decrease seen at sub-gap energies in Fig. S2b and 2c would imply S/N interfaces of reduced transparency. Note that pure tunneling behavior, i.e. vanishing conductance for sub-gap energies, is observed in none of the measured devices. During experiments, very different transparency could be obtained on the same YBCO/Au film. Devices showing low-transparent interfaces behavior as in Fig. S2b and 2c showed weaker gating effects, suggesting that in these the Au/graphene interface (instead of the superconducting graphene/normal graphene interface) is governing the junction conductance.

We could not quantify the Au/graphene contact *resistivity* characteristic of each of those behaviors. Indeed, while low-transparency behavior samples often display lower conductance than high-transparency one [compare e.g. Figs. S2a and S2c], in some cases close values of the conductance lead to opposite behaviors [compare e.g. Figs. S2a and 2b]. This suggests that the effective contact geometry changes from sample to sample. For this reason, we cannot reliably estimate the Au/graphene contact *resistivity* from our experiments.

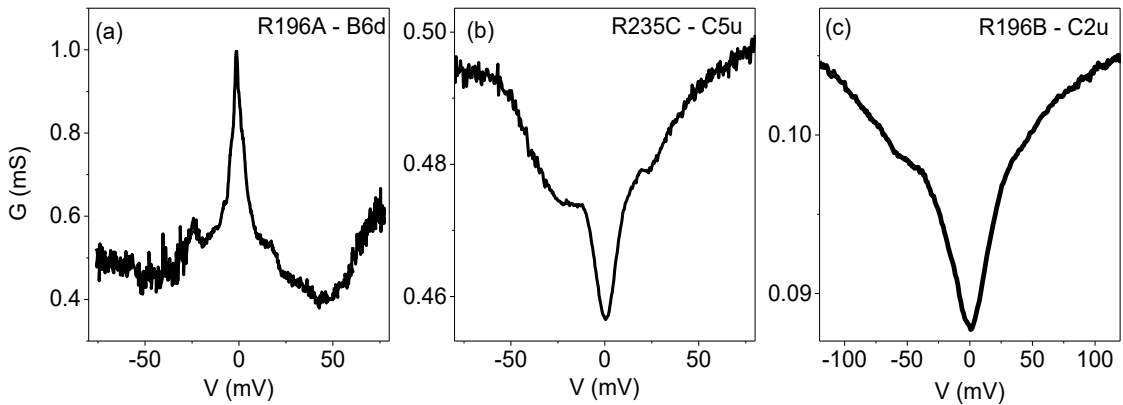

**Figure S2:** Low temperature (3 K) measurement of the conductance for different devices

## 3. Gate and graphene characterization via Hall experiments

We fabricated a graphene Hall bar in order to measure the gate capacitance and the native doping of the CVD graphene. This device was fabricated using the same type of 50 nm thick YBCO films grown on 500 μm thick STO substrates as those used for the S/N'/N junctions studied in the paper. The same oxygen irradiation procedure as for the junctions was used to turn YBCO insulating (50 nm). Then graphene was transferred on top of the insulating YBCO, and optical lithography was used both to shape the graphene into a Hall bar and to deposit gold electrodes. Measurements of $R_{xy}$ as a function of the magnetic field H and gate voltage $V_G$ were performed in a cryostat at low temperature (4K). For $V_G=0$ we obtained the native carrier concentration in graphene $n = \frac{1}{e}\frac{\partial B}{\partial R_{xy}} \sim 10^{13} cm^{-2}$.

Hall measurements were done as a several gate voltages to obtain $n(V_G)$, which is shown in Fig. S3. Note that the carrier modulation is linear with respect to the applied gate voltage. This leads to $C = \frac{\partial n}{\partial V_G} = 2\ 10^{10} cm^{-2} V^{-1}$. This is as expected from a rough estimation based on the electrical susceptibilities of irradiated YBCO $\varepsilon_{YBCO}\sim 1$ and SrTiO3 substrate $\varepsilon_{STO}\sim 10^4$ and their thickness $d_{STO} = 500$ μm and $d_{YBCO} = 50$ nm, which yields $\frac{\delta n}{\delta V_G} = \varepsilon_0 \left[\left(\frac{d_{STO}}{\varepsilon_{STO}} + \frac{d_{YBCO}}{\varepsilon_{YBCO}}\right) 2e\right]^{-1} \sim 5\ 10^{10}\ (cm^{-2}V^{-1})$.

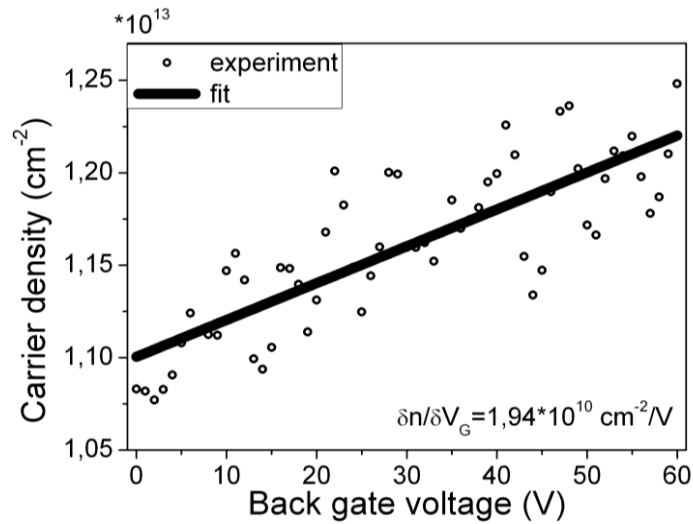

**Figure S3:** Carrier density as a function of gate voltage, the carrier density is deduced from $R_{xy}(H)$

## 4. Numerical simulations

To calculate the conductance across the S-N'-N junction depicted in Fig. 3 (g), we performed numerical simulations using the model developed by Linder and Sudbo [5]. Here we briefly recall the model and discuss the approach used to numerically reproduce the experimental results.

The Eq. (20) of [2] relates the differential conductance $g = dI/dV$ across a graphene S/N'/N junction to the probabilities of Andreev and normal reflection, respectively $|r_A|^2$ and $|r|^2$, as follows:

$$g(eV/\Delta, \chi, E_f/\Delta, E_f'/\Delta, \alpha) = \int_{-\pi/2}^{\pi/2} d\theta \left( \cos(\theta) \left(1 - |r(eV/\Delta, \theta, \chi, E_f/\Delta, E_f'/\Delta, \alpha)|^2\right) + \cos(\theta_A) |r_A(eV/\Delta, \theta, \chi, E_f/\Delta, E_f'/\Delta, \alpha)|^2 \right) \quad \text{(Eq. 1)}$$

where $\theta$ is the angle between the incoming electron wave-vector and the interface, $E_F$ the Fermi energy in normal graphene (N), $E_F'$ the Fermi energy in superconducting graphene (S), $\alpha$ the angle between the d-wave nodes of the superconducting order parameter and the S-N' interface, $\Delta$ the superconducting energy-gap, and $\chi$ the phase acquired by an electron while crossing the intermediate region N'. The probabilities are calculated by matching the wave functions at the interfaces, their expressions are given in the appendix of [5]. The simulation is performed in the thin barrier limit, in which the phase is given by [5]

$$\chi = (U_0 - E_F)w/\hbar v_F \quad \text{(Eq. 2)}$$

where $w$ is the width of the N' region, $v_F \sim 10^6$ m s$^{-1}$ and $U_0$ the barrier height.

In order to find the best fit between theory an experiments, a number of simulations of $g(eV/\Delta, \chi)$ as those shown in Fig. 3 (d)-(f) were done. Note that, for the simulations, the only free parameters were $\alpha$ and $E_F'$. All of the other parameters are known from direct inspection of the experimental conductance curves and from the Hall measurements described above. In particular:

- **$E_F$ and its dependence on $V_G$** are obtained from Hall measurements. These provide the native graphene carrier density $\sim 10^{13}$ cm$^{-2}$, and demonstrate that doping is directly proportional to the $V_G$, with $\delta n/\delta V_G = 2\,10^{10}$ cm$^{-2}$ V$^{-1}$. From that, and given $E_F = \hbar v_F k = \hbar v_F \sqrt{\pi n}$ [6], we obtain $E_F(V_G)$ –black symbols in Fig. S4. One sees that, in a very good approximation, $E_F$ is proportional to $V_G$ (red curve) within the experimental window (dashed rectangle), with $\delta E_F/\delta V_G \sim 4\,10^{-4} e$.

- **The width of the N' region, $w$.** From the above, and considering Eq. 2, it follows that the phase shift $\chi$ must be proportional to $V_G$. *Note that this prediction is consistent with the periodic modulation by $V_G$ observed in conductance measurements [see Fig. 3 (b)].* Thus, from simple inspection of Fig. 3(b), we obtain the relationship between the phase shift $\chi$ and $V_G$, $E_F$ and $n$, which is represented by the red line in Fig. S4. From the proportionality factor $\delta \chi/\delta E_F \sim 3\pi/100$ meV$^{-1}$ and Eq. 2, we estimate $w = \hbar v_F \delta \chi/\delta E_F \sim 60$ nm.

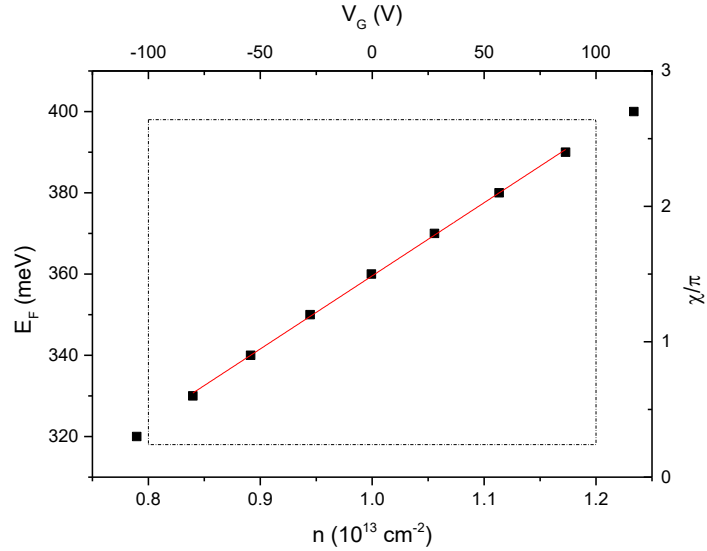

**Figure S4:** Relationship between $E_F$, $V_G$, $n$ and $\chi$

- **The superconducting gap $\Delta \sim 20$ mV** is known from the position of the conductance dips around zero-bias. That is in good agreement with expectations for YBCO along the c-axis [7].

Based on the above, we performed simulations $g(eV, \chi)$ by independently varying $\alpha$ and $E_F'$ to find the best fit to the experiments, which corresponds to $\alpha = \pi/4$, $E_F' \sim 20\Delta \sim 400$ meV.

To illustrate how sensitive the simulations are to those parameters, we show in Fig. S5 a few examples of simulations made with $\alpha$ and $E_F'$ that are off the right values.

- In the case of $\alpha$, significant departures from $\alpha = \pi/4$ lead to behaviours that radically differ from the experimental results. For instance, for $\alpha = 0$ (Fig. S5a) the conductance around V=0 shows strong "dips" around $\chi = \pi/2, 3\pi/2$ ... which are not observed in the experiments. Also, the conductance minima around $V = \Delta$ are much shallower in that simulation than in the experiments, which is also the case for $\alpha = \pi/6$ (Fig. S5b).

- The case of too high $E_f' = 30\Delta$ is shown in Fig. S5c. In this case, the depth conductance minima around $V = \Delta$ are much shallower than in the experiments.

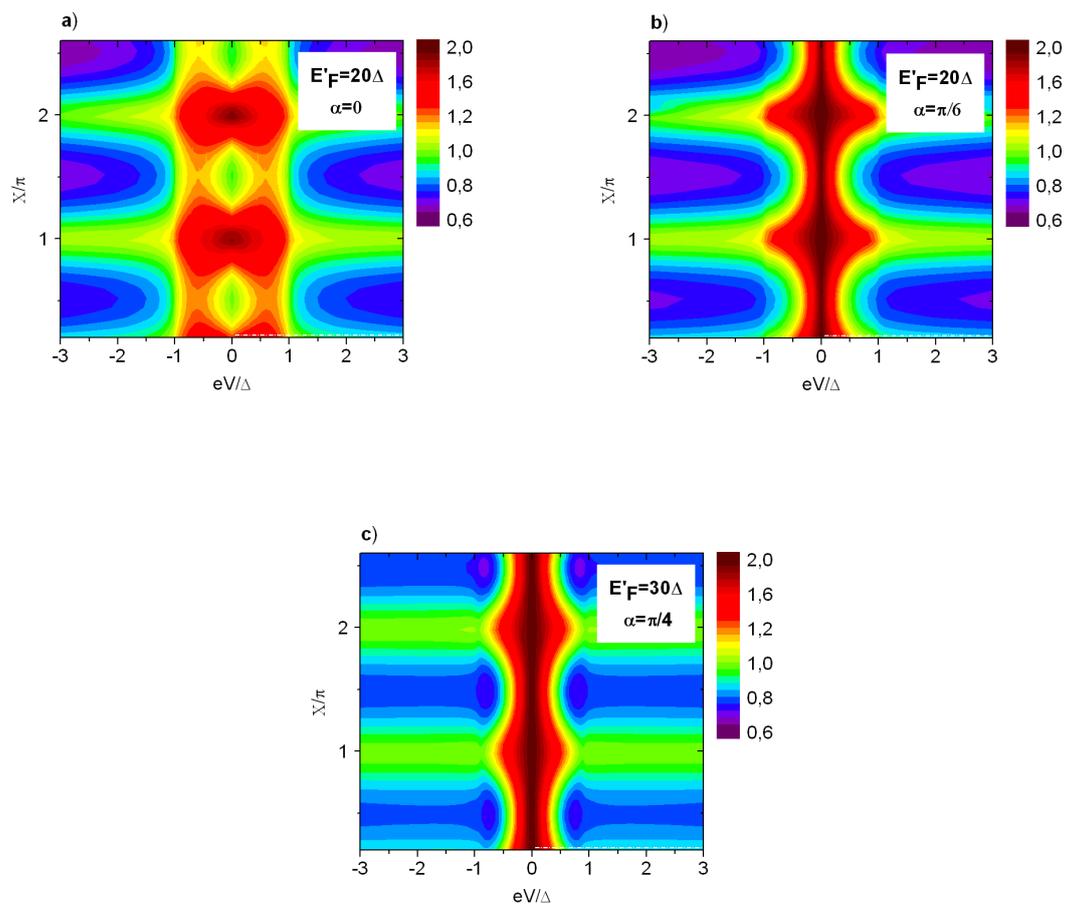

**Figure S5:** Differential conductance (color scale) as a function of bias and phase for different set of parameters (see legends)

## 5. Comparison between theory and experiments for energy $eV>\Delta$.

As discussed in the main text, the agreement between theory and experiments is best in the energy range $eV<\sim 1.5\Delta$. We show in Fig. S6 below the experimental data up to $eV=3\Delta$ (which is the upper limit of our experimental window).

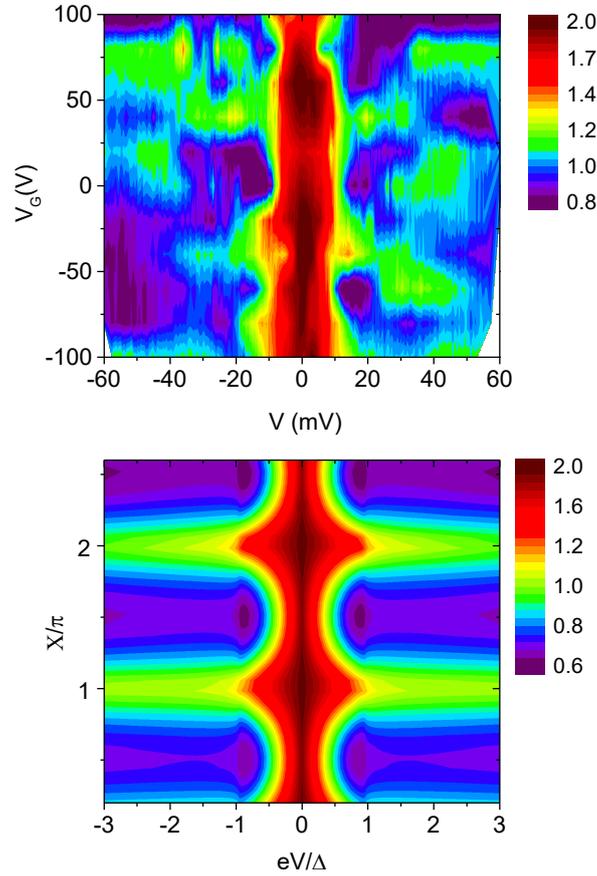

**Figure S6 :** (upper) Contour plot of the normalized conductance (color scale) as a function of junction bias V and gate voltage $V_G$. This plot is constructed from a series of measurements as those shown in Figure 2, with $\Delta V_G$=20 V. The lower graph shows numerical simulations correlative of the measurements. Notice that in the simulation the bias V is normalized by the superconducting gap $\Delta$, and the experimental quantity $V_G$ is replaced by the phase $\chi$.

The disagreements between theory and experiments, which appear when $eV>\sim 1.5\Delta$, suggest that within this energy range some of the model approximations are probably not well suited to our experiments. For instance, for high enough V the assumption that the energy barrier is "square" may no longer be valid (as it increases, V gradually becomes non-negligible with respect to the difference between $E_F$ and $E_F$'), and thus a "trapezoidal" barrier should be considered. Also, when $eV>\Delta$ the current across the junction is primarily carried by normal electrons (instead of Andreev pairs), so that coherence (and thus interference effects) are probably preserved over a shorterlength scale. Those effects, among others, should be theoretically taken into account for a better match between theory and our experiments in the range $eV>>\Delta$.

## 6. Origin of N' and energy barrier $U_0$.

To understand the origin of the barrier $U_0$, we start from the hypothesis that graphene doping is different i) on top of Au (region S), ii) on top of insulating YBCO (region N), and iii) in an intermediate region N' in between S and N. Following theory and experiments, particularly work related to Klein tunneling in graphene [8], that situation implies a potential step for electron travelling from S to N across N'.

Since Au and (insulating) YBCO are very different, it is straightforward that doping is different in "S" and "N". Indeed graphene doping is strongly dependent on the electronic and structural properties of the underlying substrate [9] and hence different doping in "S" and "N" is expected.

The reason why a region N' exists with different doping as in N is not as straightforward. Various non-exclusive explanations are plausible.

The most plausible one is that the cross-section of the device can be viewed as two contiguous capacitors having the same lower plate (back of the STO substrate covered with silver paste) but different top plates and dielectric media (see Fig. 2c). The first capacitor (left) has superconducting YBCO/Au as top plate and STO as the dielectric medium. At variance, in the second capacitor (right) graphene is the top plate, and the dielectric media are STO in series with irradiated YBCO. In this configuration, strong fringe field exists at the boundary between both capacitors (which corresponds to the S/N interface, see Fig. 2c). Because of the fringe fields, the excess charge in the S/N interface is locally different from that induced far from it. We performed calculations using finite element software (COMSOL) from which we found that the excess charge strongly decays within the first 100 nm from the S/N interface. This effect supports the existence of an intermediate N' region at the S/N interface.

In addition to the above, the concomitance of two effects cannot be excluded:

1) The 4 nm Au layer covering superconducting YBCO forms a step with respect to uncovered, irradiated YBCO. Thus, one expects graphene to be locally corrugated and possibly "suspended", which should lead to different local doping in that region and further from the step.

The electronic properties of the irradiated YBCO are different within few nm of the Au step and further from it, since the boundary between irradiated and unirradiated YBCO is not sharp, but has a ~ 10 nm scale characteristic size. This may lead to different local graphene doping within that region between irradiated and unirradiated YBCO.

## 7. Analysis of the oscillatory dependence of the conductance on the gate voltage.

Figure S7 shows g(V$_G$) for different V, in the energy range eV ≤ Δ. Note that, since g(V) measurements are current biased –V is measured– the same exact V is not available for every V$_G$. To circumvent this limitation, add statistical weight to the data, limit noise and provide an indication of the measurement uncertainty, we plot for each V$_G$ all the points within ±1mV around the nominal V. The spline line that interpolates the data goes through the average value of these points.

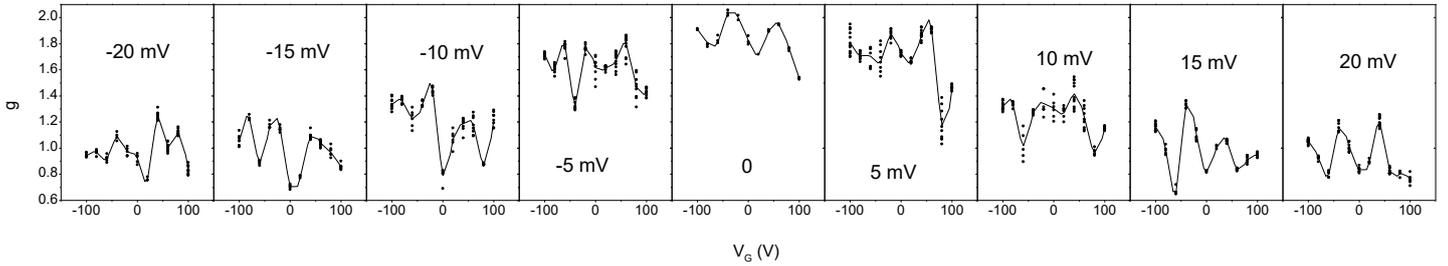

**Figure S7 :** Conductance vs. gate voltage V$_G$ for different V (see labels, the uncertainty is ±1mV), which correspond to vertical "cuts" of the Fig. 3 (a).

In Figure S7, one can see that the conductance oscillates as a function of V$_G$. In order to quantitatively analyze the oscillatory behavior, Fig. S8 shows the fast Fourier transform (FFT) of all the measured data points [the entire series of g(V,V$_G$)]. For comparison, we show two FFT: one (Fig. S8a) with a cut-off at V< 30 mV ~ 1.5 Δ (this corresponds to the energy range in which theory and experiments best agree) and one for the entire experimental window (Fig. S8b). The FFT are shown in a color code on the x-y plane of the figures, and as a histogram on the x-z and y-z projections. We can see in both figures that the FFT unveils a well-defined frequency in the gate-voltage domain, pointed by the red arrows.

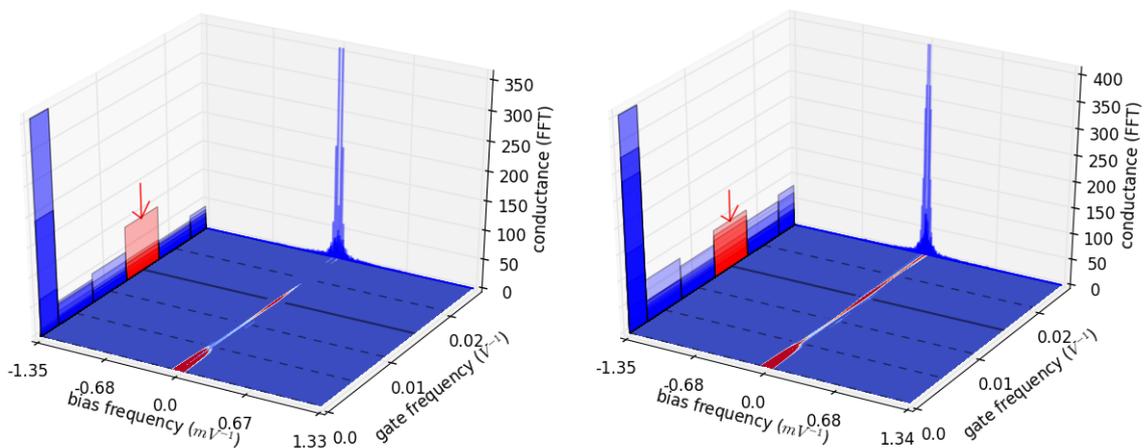

**Figure S8** : (a) FFT of the g(V,V$_G$)] for V<30 mV (b) for V<60 mV.

**REFERENCES**

[1]  C. Visani, Z. Sefrioui, J. Tornos, C. Leon, J. Briatico, M. Bibes, A. Barthélémy, J. Santamaría, and J. E. Villegas, Nat. Phys. **8**, 539 (2012).

[2]  A. Sharoni, I. Asulin, G. Koren, and O. Millo, Phys. Rev. Lett. **92**, 17003 (2004).

[3]  A. Volkov and K. Efetov, Phys. Rev. Lett. **102**, 77002 (2009).

[4]  E. Scheer, W. Belzig, Y. Naveh, M. H. Devoret, D. Esteve, and C. Urbina, Phys. Rev. Lett. **86**, 284 (2001).

[5]  J. Linder and A. Sudbø, Phys. Rev. B **77**, 64507 (2008).

[6]  A. H. C. Neto, N. M. R. Peres, K. S. Novoselov, and A. K. Geim, Rev. Mod. Phys. **81**, 109 (2009).

[7]  S. Kashiwaya, Y. Tanaka, M. Koyanagi, H. Takashima, and K. Kajimura, Phys. Rev. B **51**, 1350 (1995).

[8]  C. W. J. Beenakker, Rev. Mod. Phys. **80**, 1337 (2008).

[9]  Q. H. Wang, Z. Jin, K. K. Kim, A. J. Hilmer, G. L. C. Paulus, C. Shih, M. Ham, J. D. Sanchez-yamagishi, K. Watanabe, T. Taniguchi, J. Kong, P. Jarillo-herrero, and M. S. Strano, Nat. Chem. **4**, 724 (2012).